\documentstyle[aas2pp4]{article}  

\newcommand{\beq}{\begin{equation}}
\newcommand{\eeq}{\end{equation}}
 
\begin{document}
\twocolumn
\title{\bf Massive Cooling Flow Clusters Inhabit Crowded Environments} 
 
\author{Chris Loken$^1$, Adrian L. Melott$^2$, Christopher J. Miller$^3$} 
 
\vskip 0.2in 
 
\baselineskip 10pt 

\affil{$^1$Dept. of Physics \& Astronomy, University of
 Missouri, Columbia, MO 65211}
\affil{$^2$Dept. of Physics \& Astronomy, University of Kansas,
Lawrence, KS 66045
} 
\affil{$^3$Dept. of Physics \& Astronomy, University of Maine,
Orono, ME 04469}
 
\vskip 0.5in

\begin{abstract} 
With the availability of large-scale redshift survey data, it is now 
becoming possible to explore correlations between large-scale structure
and the properties and morphologies of galaxy clusters. 
We investigate the spatial distributions of a 98\% complete, volume-limited
sample of nearby ($z<0.1$) Abell clusters with well-determined
redshifts and find that cooling flow clusters with high mass-accretion 
rates have nearest neighbors which are much closer than those of other 
clusters in the sample (at the 99.8\% confidence level),
and reside in more crowded environments out to 60 $h_{50}^{-1}$ Mpc. 
Several possible explanations of this effect are discussed.
\end{abstract}


\section{Introduction}

Intriguing correlations between cluster properties and
larger-scale structures have been noted over the past two decades. For example,
the major axes of Abell clusters may be aligned with nearby
clusters (e.g., Bingelli 1982; Plionis 1994),  X-ray substructure within clusters appears
to point to neighboring Abell clusters (West, Jones, \& Forman 1995),
and the tails of Wide-Angle Tailed (WAT) radio sources within clusters 
tend to be aligned with the long axis of the nearest supercluster
(Novikov et al.~1999). All of these alignment effects are
broadly consistent with the scenario of hierarchical structure
formation whereby clusters form through the flow of material along
sheets and filaments (e.g., Shandarin \& Klypin 1984; Pauls \& Melott 1995; 
Bond, Kofman, \& Pogosyan 1996;
Colberg et al.~1999). Here we investigate another possible
link between cluster properties and larger-scale structures:
the environments within which cooling flow clusters
are located.

Cluster cooling flows are believed to result when gas 
within the central $\sim100$ kpc cools radiatively on a time-scale less 
than a Hubble time; the resultant decrease in pressure allows gas
to flow towards the core (see, e.g. the review of Fabian 1994). 
Several groups have estimated that $(60-90)\%$ of clusters contain
cooling flows (Edge, Stewart, \& Fabian 1992; White, Jones \& Forman 1997;
Peres et al.~1998) with mass deposition rates 
(obtained by detailed deprojection analysis of their X-ray images)
ranging from a few to more than one thousand 
M$_\odot$ yr$^{-1}$.
The environmental factors which determine whether or not a given cluster 
will develop a cooling flow are still unclear and have not been
properly addressed by  numerical simulations.
However, the development of a strong cooling flow was 
observed in the single cluster simulation of Katz \& White (1993), and
detailed 1D simulations incorporating many realistic effects 
suggest that strong cooling flows are a natural outcome of
cluster formation over a wide range in masses (Knight \& Ponman 1997).

Though cooling flows may be a generic feature of cluster formation,
it appears that the
recent merger history of a cluster is closely linked to the 
existence and/or strength of a cooling flow. McGlynn \& Fabian (1984) and 
Fabian \& Daines (1991) suggested that cooling flows may be disrupted
by cluster mergers. This is supported by the fact that there
are no cooling flows in several massive clusters which appear to have
undergone recent mergers (e.g., Coma:  Burns et al.~1994; A2255: Burns et al.~1995).
Unfortunately, the observational picture is not entirely clear as there 
are also cooling flow clusters which appear to have
undergone a merger (e.g., A1664: Allen et al.~1996; A2597: Sarazin
et al.~1995). Nonetheless, the observations are 
consistent with a scenario in which recent and/or massive mergers
tend to disrupt any pre-existing cooling flow. For example,
Buote \& Tsai (1996) demonstrated that cooling flow strength is 
anti-correlated with the degree of X-ray substructure (which is likely
a measure of recent merger activity).
Recent numerical simulations (Burns et al.~1997; Gomez et al.~1999) of 
cooling flow mergers
indicate that high-$\dot{M}$ flows are more difficult to disrupt
than low-$\dot{M}$ flows and that the likelihood of disrupting 
a given cooling flow increases with the mass of the infalling cluster.

To date, there has been no systematic investigation of the
large-scale environments within which cooling flow 
clusters are found. A possible connection between large-scale structure 
and cluster cooling flows comes from
the evidence that mergers can  disrupt cooling flows which, in turn,
suggests that massive cooling flows have been
relatively undisturbed for some time.
As Abell clusters are thought to delineate the supercluster
distribution rather well (e.g. Bahcall et al. 1994), we
decided to investigate the spatial distribution of Abell 
clusters around cooling flows.

\section{Procedure}

\subsection{The Abell Cluster Sample (AC)}

We began by constructing a complete sample of 
all rich Abell/ACO clusters (Abell 1958; Abell, Corwin,
\& Olowin 1989) with
$R \ge 1$, m$_{10} \le 17.0$, $|b| \ge 30^{\circ}$, and 
spectroscopically-determined redshifts 
$0.012 \le z \le 0.10$ (where $R$ is Abell's richness class  
and m$_{10}$ is the magnitude of the tenth brightest galaxy within the cluster).
The cut in galactic latitude rejects regions of the sky where
obscuration and poor sampling may plague the ACO catalog while
our magnitude limit corresponds to $z \sim 0.13$ assuming an
m$_{10}$-$z$ relationship (e.g., Batuski \& Burns 1985).
The total number of ACO clusters satisfying these cuts is 284, of which 
we kept the 277 with  measured redshifts.  The majority of the cluster redshifts
come from the ESO Nearby Abell Cluster Survey (Katgert et al.~1996) 
and the MX Northern Abell Cluster Survey (Slinglend 
et al.~1998), although $\sim$ 50 are unpublished 
redshifts from Miller et al.~(1999b).

It is important to note the level of completeness (98\%) and quality of this
Abell/ACO dataset. Miller et al.~(1999a) and Peacock \& West (1992)
have shown that $R \ge 1$
Abell/ACO cluster suffer much less from spurious cluster selection (due to
line-of-sight anisotropies) than do samples containing $R= 0$
clusters. By limiting the samples to $R \ge 1$ clusters we are sacrificing
some poorer clusters with published mass deposition rates.
However, since we are interested in nearest neighbors, it is imperative 
that our base  sample
be significantly complete in redshift measurements. A magnitude-limited and volume-limited subset of $R \ge 1$ Abell/ACO 
clusters is the only viable dataset for an analysis such as this.
This sample of Abell/ACO clusters has a nearly constant spatial
number density ($\bar{n} = 9 \times 10^{-7}h_{50}^{3}$Mpc$^{-3}$)
out to $z \sim 0.10$.

After selecting an initial sample based on the above criteria, we removed
any clusters which were closer to a boundary of our volume than to their
nearest neighbor in order to carry out the nearest-neighbor analysis
described below. 
The remaining set of 202 clusters, which shall be referred to as AC,
is a 98\% complete, volume-limited sample of Abell clusters.

\begin{table}
\begin{tabular}{clc} \hline
Cluster & $\dot{M} (M_\odot$yr$^{-1})$ & Ref \\ \hline
A0085& $      198_{-      52}^{+      53}$ & P \\
A0401& $      42_{-      42}^{+      82}$ & P \\
A0970& $      20_{-      20}^{+      32}$ & W \\
A0978 &       500&          S \\
A1126 &       500&          S \\
A1644& $      11_{-      5}^{+      40}$ & P \\
A1650& $      280_{-      89}^{+      464}$ & P \\
A1651& $      138_{-      41}^{+      48}$ & P \\
A1795& $      381_{-      23}^{+      41}$ & P \\
A1837& $      12_{-      12}^{+      29}$ & W \\
A1983& $      6_{-      6}^{+      10.8}$ & W \\
A1991& $      37_{-      11}^{+      36}$ & W \\
A2029& $      556_{-      73}^{+      44}$ & W \\
A2063& $      37_{-      12}^{+      7}$ & P \\
A2065& $      13_{-      6}^{+      14}$ & P \\
A2107& $      7.1_{-      7.1}^{+      17.2}$ & W \\
A2151& $      166_{-      41}^{+      51}$ & W \\
A2152& $      20_{-      20}^{+      13}$ & W \\
A2197& $      2.4_{-      2.4}^{+      3}$ & W \\
A2199& $      154_{-      8}^{+      18}$ & P \\
A2556& $      81_{-      81}^{+      105}$ & W \\
A2657& $      44_{-      24}^{+      36}$ & W \\
A2670& $      41_{-      41}^{+      71}$ & W \\
A3112& $      376_{-      61}^{+      80}$ & P \\
A3158& $      25_{-      25}^{+      74}$ & P \\
A4059& $      130_{-      21}^{+      27}$ & P \\
\\
\hline
\end{tabular}
\caption{The CF (cooling flow) sample of clusters. These are all the AC 
clusters with non-zero published values of $\dot{M}$. The
references for the quoted values are: P (Peres et al.~1998),
S (Sarazin 1986), and W (WJF97). The quoted values assume $H_o$=50
km/s/Mpc. 
}
\label{mdot_table}
\end{table}

\subsection{The Cooling Flow Clusters}

Using NED, we searched the astronomical literature for papers
dealing with AC clusters in order to find
published determinations of $\dot{M}$. The single most comprehensive 
source of clusters 
with detailed X-ray deprojections and analysis is the 207 clusters
with {\it Einstein} data studied by White, Jones, \& Forman (1997; 
hereafter WJF). Our second major source is the analysis
of ROSAT data performed by Peres et al.~(1998) on a sample of the 
55 X-ray brightest clusters originally identified by Edge et al.~(1990). 
For clusters that appear in both the WJF and Peres
et al.~lists, we use the ROSAT-derived values of $\dot{M}$ from 
Peres et al. (taking the PSPC-derived value whenever available). 
A handful of additional $\dot{M}$ values were gleaned from other
papers (Sarazin 1986; Edge \& Stewart 1991;  Pierre \& Starck 1998). 

All told, 55 of our AC clusters have published values of
$\dot{M}$ and 26 of these have non-zero mass accretion rates
(see Table~\ref{mdot_table}).
The fact that half our clusters with estimated $\dot{M}$ have non-zero 
mass accretion rates is consistent with the cooling flow occurence rate
estimated by WJF.
For the analysis discussed below, we created several different subsamples
of cooling-flow clusters; CF1 contains only those clusters with $\dot{M}>50 M_\odot$yr$^{-1}$, 
CF2 is a sample with $\dot{M}>35 M_\odot$yr$^{-1}$, and CF3 is a sample of
those clusters with 
$\dot{M}-\sigma_{\dot{M}}>20 M_\odot$yr$^{-1}$ (where $\sigma_{\dot{M}}$ is
the estimated 1-$\sigma$ lower uncertainty).

\subsection{The Non-Cooling Flow Clusters}

The WJF catalog quotes $\dot{M}=0$ for 27 of our AC clusters. However,
after taking into account the spatial resolution of the images, WJF
conclude that only 10 of these are truly excluded as being
cooling flows. We take these to be our most secure sample 
of non-cooling flow clusters (NCF1). This sample comprises:
A0019, A1185, A1213, A1291, A1656, A1913, A2040, A2147, A3158 and A3744.
Our second sample of non-cooling flow clusters, NCF2,
contains all those AC clusters in WJF which have $\dot{M}=0$ and an
upper uncertainty less than 20. After removing A2065 which
has $\dot{M}\ne 0$ according to Peres et al.~(1998), the NCF2 sample
is made up of: A0150, A0154, A0168, A0399, A0500, A0690, A1185,
A1213, A1377, A1656, A1809, A1913, A2040, A2079, A2092, A2124, A2147,
and A3744. 

\subsection{Selection Effects}

Our AC sample is  well-suited for
the spatial distribution analyses we are interested in.
Unfortunately, the 55 AC clusters with published 
values of $\dot{M}$ form an inhomogeneous sample since they were
observed and studied by various authors for different
reasons. 
Nonetheless, we find (\S3) that subsamples of cooling flow 
and non-cooling flow clusters drawn from this sample 
of clusters with known-$\dot{M}$ differ significantly in their 
nearest-neighbor distributions. 
It is difficult to imagine how such an effect could arise spuriously
if the clusters with known-$\dot{M}$ form a random sampling
of the AC clusters. We point out that the clusters with known-$\dot{M}$
cannot be selected from AC on the basis of an X-ray flux or luminosity cut
suggesting that they are more-or-less randomly selected. 
Though all the AC clusters are potential members of the X-ray Brightest 
Abell Clusters (XBACs) catalog (Ebeling et al.~1996), only 61 AC clusters
and only 37/55 of the AC clusters with known-$\dot{M}$ meet the XBACs flux-limit
(even our two most massive flows do not make the catalog).

\section{Results}

We wished to determine if the existence of cooling flows is
influenced by the
proximity of other clusters.  We tabulated the distance
to the nearest neighbor for every cluster in our AC sample and then
considered whether the distribution of these nearest-neighbor
distances is statistically different for any of our subsamples.
Since clusters are correlated, we expect these
distances to be smaller than the $~104 h_{50}^{-1}$ Mpc mean
distance between our clusters.
Distances to all clusters were determined assuming 
$q_0 = 0$ and $H_0 = 50$ km s$^{-1}$ Mpc$^{-1}$.

Since we are looking for a systematic difference in the typical distance of
the nearest neighbor, the Wilcoxon Rank-Sum test is appropriate (Lehmann \&
D'Abrera 1998).  This test consists of putting all the nearest-neighbor
distances
in rank order.  The sum of the ranks of each subset is examined
for consistency with the null hypothesis that there is no difference
in the rankings. 
We emphasize that the significance given by this test
explicitly accounts 
for fluctuations that may arise due to sample size.

Each subsample was compared with its own control sample made up of all the 
AC clusters not in that particular subsample.
The results of the analysis are given in Table~\ref{results}, including the
significance of the result that clusters in the sample are
systematically closer to their nearest neighbor than those in their control
group (AC minus the subsample).
Although it is not used in the assessment of significance, we also
tabulate the mean distance
to the nearest-neighbor ($\overline{nnd}$) for each subsample and
its corresponding control sample. The analysis shows one
very significant result;  massive cooling flows (CF1) tend to 
have much {\it nearer} closest neighbors than their control clusters.
In fact, the average nearest-neighbor distance
($\overline{nnd}$) is just half what it is for the remainder of the
AC sample. When less massive cooling flows are included in the
subsample (CF2 and CF3), the effect becomes less pronounced 
(though still significant at the 95\% level).
The significance
is lowest for CF2 which, by definition,  includes some clusters 
whose $\dot{M}$ values are consistent with zero. We performed the
same analysis with the non-cooling flow samples to make sure that
this is not a spurious effect. Indeed, the most secure sample of
non-cooling flow clusters (NCF1) is not significantly different from 
the remainder of the AC sample. The second non-cooling flow sample (NCF2)
contains some potential cooling flows (since $\dot{M}=0^{+20}_{-0}$) and has a higher likelihood of differing
from its control sample than NCF1.
This suggests a possible trend in which $\overline{nnd}$ decreases with
increasing $\dot{M}$.

\begin{table}[t]
\begin{tabular}{rcccl} \hline
Subsample & Subsample & $S(\%)$  & $\overline{nnd}$ (h$_{50}^{-1}$Mpc)
       & $\overline{nnd}$ (h$_{50}^{-1}$Mpc)\\ 
       & size   &   &   (subsample)   & (control) \\ \hline
\\
CF1 & 12 &99.8 & 18.4  & 38.4\\
CF2 & 17 &95.5 & 28.6 & 38.0\\
CF3 & 14 &97.6 & 26.0 & 38.0\\
NCF1& 10 &80.8 & 30.2 & 37.6\\
NCF2& 18 &89.1 & 29.2 & 38.0\\
\hline
\end{tabular}
\caption{Results of the Wilcoxon Rank-Sum test comparing nearest-neighbor
distances ($nnd$) of our various subsamples with the rest
of the AC sample. 
S is the significance (\%) that the subsample is systematically more
crowded than the control.
The average nearest-neighbor distance ($\overline{nnd}$) for both 
the subsamples and the control samples are also
tabulated. }
\label{results}
\end{table}

\begin{figure}[t]
\plotone{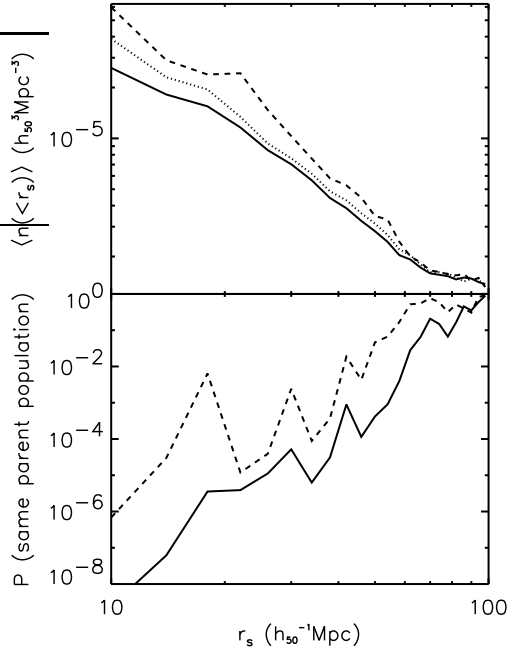}
\caption{
The top panel shows the mean number density of galaxy clusters in spheres 
of specified  radii centered on the clusters in: 
the AC sample (solid line), the CF1 sample (dashed line), and
the proportioned AC sample (dotted). More significantly, the bottom panel
shows the probability that the densities around CF1 clusters
(whose mean is shown 
in the upper panel) could result from
a random selection of the AC sample (solid line)
or the proportioned AC sample (dashed line).}
\label{density}
\end{figure}

As a double-check, we calculated the 
mean number density of clusters in spheres of specified radii centered
on the clusters of the entire AC sample and on those in the
CF1 sample. For any given radius $r$, only those AC (or CF1) clusters
which were at least a distance $r$ from a boundary of the AC sample
volume were considered. We used all 277 ACO
clusters to find density inside a given radius.
The results are depicted in the top panel of Figure~\ref{density} and
show that local mean densities (for $r \leq 60$h$_{50}^{-1}$Mpc) are higher 
for the CF1
clusters (dashed line) than for the entire AC sample (solid line). 
The significance of the difference is assessed in the bottom panel which
shows the probability (using a Kolmogorov-Smirnov test; Lehmann \& 
D'Abrera 1998) that the {\it distribution of
densities} at a specific radius could result from a random selection
of the AC parent population. 
The differences are highly significant out to 60 h$_{50}^{-1}$ Mpc, although
roughly what would 
be contributed by an average of one extra neighbor near each CF1 cluster.

Our main result is the close nearest neighbor and generally increased density
of the large-scale environment for large
distances around massive cooling flow clusters.  One might worry that
this effect could be a richness bias as our CF1 cluster sample has a larger 
fraction of R $>$ 1 clusters than the AC sample,
causing the CF1 sample to be more strongly 
spatially correlated (Kaiser 1984) simply on this basis. After removing
the CF1 clusters, the AC sample contains 155 R=1 clusters, 33 R=2
clusters and 2 with R=3 whereas the CF1 sample has 6 R=1 and 6 R=2
clusters.
We therefore constructed four ``proportioned" subsamples of AC that 
each contained all
33 of the R=2 clusters as well as 33 randomly chosen R=1 clusters.
Averaging our Wilcoxon Rank-Sum
test for these four samples 
produced a 99.5\% confidence that the CF1 sample is more crowded than
these richness-proportioned control groups. The mean nearest neighbor 
distance for the
proportioned controls is (34-38) $h_{50}^{-1}$ Mpc as opposed to 
18.4$h_{50}^{-1}$ Mpc
for CF1.  This, along with the diagnostics
for the proportioned set in Figure 1 indicate that the CF1 sample is 
significantly different from both AC and a richness-proportioned 
subsample of AC. 
We conclude that any richness  bias in our result is a small effect.

\section{Conclusions}

We find that the most massive cooling flows (those with 
$\dot{M}>50 M_\odot$yr$^{-1}$)
have significantly closer nearest-neighbors 
($\overline{nnd}=18.4 h_{50}^{-1}$ Mpc)
than the typical cluster ($\overline{nnd}$=38.4$h_{50}^{-1}$ Mpc) in our
complete, volume-limited sample of rich Abell clusters.
We cannot exclude that this result is related to some unknown
selection effect connected with why the clusters were originally selected
for X-ray observation. However, this seems unlikely given that Table 2
shows a trend for the significance of the result to decrease as less 
massive cooling flows are added to the subsample and to disappear
completely when samples of non-cooling flows are used.

How can one interpret our result that massive cooling flow clusters tend
to reside in environments that are
crowded with other clusters?
One might expect
that mergers would be much more common in such regions and that therefore
cooling flows would be {\it less} likely to survive (e.g.~Fabian 1994).
However, it is possible that these regions are crowded because
they have not yet finished collapsing and their member clusters have
not recently experienced violent mergers. By this interpretation, many
clusters which appear isolated may have recently gobbled up their
neighbors and should show signs of dynamical activity.

Since cooling times will decrease as density increases (at fixed
temperature), we might expect to find cooling flows in regions of 
high  baryon density which would typically  be
regions of high matter and cluster density 
(Bardeen et al.~1986).
However, cooling flows are found not
only in clusters, but also in anemic small groups and individual galaxies.

Cooling flow clusters are usually highly symmetric indicating 
dynamical relaxation.
Thus one might expect cooling flow clusters to be old clusters,
formed from high-amplitude perturbations which collapsed early.
Especially with the kind of cosmological power spectra that appear viable at
this time, one would expect a high-amplitude peak to be surrounded by a 
larger region of high density.

We note that the density of clusters around a cooling flow cluster 
is significantly
higher out to about 60 $h_{50}^{-1}$Mpc.  This is interestingly close to
the lengthscale for supercluster effects on WATs as found in
Novikov et al.~(1999) and to the wavelength scale going nonlinear today.
It is suggestive of a general link between cooling flows and the current
collapse of larger-scale perturbations. It may be related to a coherent
gas flow into the clusters on an even larger scale
than previously suspected.

\medskip

\noindent 
{\bf Acknowledgments}
This work was supported by NSF grant AST-9896039 (CL),
NSF-EPSCoR (ALM), and NASA-EPSCoR through the
Maine Science and Technology Foundation (CM).
We thank Keith Ashman, Jack Burns, David Batuski, Bob Nichol, and 
Kurt Roettiger for useful discussions.

\end{document}